**Topological Plasmonic Ring Resonator**

*Fatemeh Davoodi[1,*] and Nahid Talebi[1,2,*]*

[1]*Institute of Experimental and Applied Physics, Kiel University, 24098 Kiel, Germany*

[2]*Kiel Nano, Surface and Interface Science KiNSIS, Christian Albrechts University, Kiel, Germany*

**Abstract:**

Topological plasmonic provides a new insight for the manipulation of light. Analogous to exotic nature of topological edge states in topological photonics, topological plasmonic combines concepts from topology and plasmonics. By utilizing topological protection, plasmons can be made to propagate without significant scattering or decay, even in the presence of defects or disorder. Herein, we present a study on the design, characterization, and manipulation of topological plasmonic chains of discs based on the Su-Schrieffer-Heeger model made into a ring resonator. The investigation focuses on exploring the unique properties of these resonators and their potential to support topologically protected edge modes, within the continuum of rotationally symmetric optical modes of a ring. To observe the topological edge modes in rotationally symmetric chains, we employ a symmetry-breaking excitation technique based on electron beams. It analyzes the influence of parameters such as dimerization and loop numbers on the presence of topological modes. Additionally, we explore the manipulation of electron impact positions to control the direction of propagation and selectively excite specific bulk or edge modes. The findings contribute to a deeper understanding of topological effects by numerically investigating their design principles and exploring techniques to manipulate topological edge modes. The insights gained from this study have implications for the development of nanoscale plasmonic systems with customized functionalities, potentially impacting areas such as nanophotonics and quantum information processing.

**Introduction:**

In recent years, topological plasmonics has emerged as a new frontier in manipulating light at the nanoscale. Topological systems, characterized by intriguing nontrivial properties that are robust against perturbations, have captivated the attention of researchers across diverse disciplines.[1-4] Analogous to their electronic counterparts, topological plasmonic systems[5,6] exhibit remarkable phenomena, such as protected edge states, rendering them highly promising for novel applications in, nanophotonics[7,8] quantum information processing,[9,10] and beyond.[11-13] One of the pioneering topological systems in the realm of plasmonics has been realized by the one-dimensional Su-Schrieffer-Heeger (SSH) model.[14-17] This model, originally introduced to describe the behavior of electrons in polyacetylene chains, has been adapted to the realm of plasmonic nanostructures.[18-20] In the context of plasmonic SSH nanochains, a chain of metallic nanoparticles with alternating spacing serves as the ideal platform to explore and harness topological properties. By controlling the interplay between the nanoparticle interactions via the dimerization and periodicity within these chains, the system unveils a plethora of intriguing phenomena.

Here, we explore the properties of topological plasmonic chains, with a particular focus on the mesoscopic analogous of SSH rings. We investigate the impact of the coupling efficiency on the energy band spectrum and the emergence of topological nontrivial modes within these rings. These structures, formed by interconnected gold nanoparticles, enable a robust plasmonic control.

We first analyze the tight-binding Hamiltonian of the SSH model, which uses the coupling strengths between intra- and interunit. The coupling strengths are obtained by modeling the responses with the hybridization theory, unveiling in-phase and out-of-phase bonding and antibonding states. The delicate balance between these modes defines the topological phase of the system, leading us to explore the critical parameter regions that enable topological nontriviality. One dimensional SSH lattices have been investigated extensively elsewhere.[11,16] In this paper, we explore the effect of the nano-particle interactions controlled by the dimerization and periodicity on the response of a plasmonic ring resonator formed by a topological plasmonic chain.

Intriguingly, the nanoparticle coupling strengths and the ring's geometric configuration allow us to uncover topological edge states that are formed by symmetry-breaking excitation. By selectively exciting specific sites within the SSH chain ring, we gain exquisite control over the propagation direction, shedding light on the profound connection between the electromagnetic response and the winding number, a hallmark of topological phenomena. To unveil the energy band structure and eigenmodes of these plasmonic systems, we employ numerical simulations. Our findings open up new possibilities and sparks groundbreaking applications for plasmonic devices and topological photonics.

**Topological Plasmonic Chain**

The non-trivial topological plasmonic system introduced here is as an analogue of the one-dimensional SSH model, comprising a chain of metallic nanoparticles arranged with alternating spacing, as illustrated in Figure 1(a). Each nanoparticle possesses a diameter ($d$), height ($h$), and unit cell length ($a$), while the distance between the centers of two sublattice particles is denoted by $d_1$ and $d_2$. The parameters $d_1$ and $d_2$ act as tuning parameters for the interaction strength ($v$ and $w$) between the nanoparticles. To investigate the interplay between nanoparticle interaction, dimerization, and periodicity in a plasmonic ring, we consider the configuration demonstrated in Figure 1(b). By employing nanoparticle substitution, we consider two kinds of nearest-neighbor hopping interactions ($v$ and $w$), similar to those used in the one-dimensional SSH model.[21] Depending on the specific arrangement of nanoparticle interactions, two configurations are obtained, referred to as 'trivial' and 'topological' SSH rings. To determine the topological behavior of the SSH ring system, we numerically compute the energy eigenvalues of the ring as a function of the angle ($\theta$) between nanoparticles, which serves as the dimerization parameter (Figure 1(c)). The ring consists of 16 unit cells, where each unit cell comprises two different nanoparticles. The key findings from our detailed numerical analysis are as follows: (i) The energy band structure is significantly modified compared to a one-dimensional SSH plasmonic chain. (ii) Based on the two kinds of unit cells in the SSH ring, two topologically distinct phases are observed for the cases of $v > w$ (trivial phase) and $v < w$ (topologically nontrivial phase). The distinction can be directly determined by the topological invariant, namely the winding number ($\zeta$) of phase $\varphi$ across the Brillouin zone:[22]

$$\zeta = \frac{1}{2\pi} \int_{BZ} \frac{\partial \phi(k)}{\partial k} dk \qquad (1)$$

$\zeta > 1$ for the topologically nontrivial phase ($v < w$), and $\zeta = 0$ for the trivial phase ($v > w$). The winding number represents the number of times that the phase $\varphi$ winds around the circle as $\theta$ traverses a complete Brillouin zone, which has a direct relation with the loop number $M$ in the SSH ring:

$$E(\phi(k) + \delta\phi(k)) = e^{iM\delta\phi(k)} E(\phi(k)) \qquad M = 0, 1, \ldots, N-1 \qquad (2)$$

where $\delta\phi(k) = \dfrac{2\pi}{2N}$ is the phase shift between two particles and $N = 16$ is the number of unit cells.

(3) Band structure gaps appear in both trivial and topologically nontrivial modes, with the most notable difference being the existence of two topological states within the topological gap in the latter case. These states are observed in both right-handed and left-handed circular polarization and can be associated with the winding direction (see Figure S1 for more details). The fact that the states in the gap are attributed to topological states, despite the absence of any structural edge, will be later clarified. To conduct a comprehensive investigation of the topological SSH chain ring based on the tight-binding method, the nanoparticle coupling strength is defined as the hopping strength. We hence take a step back and first analyze a linear chain of nanoparticles.

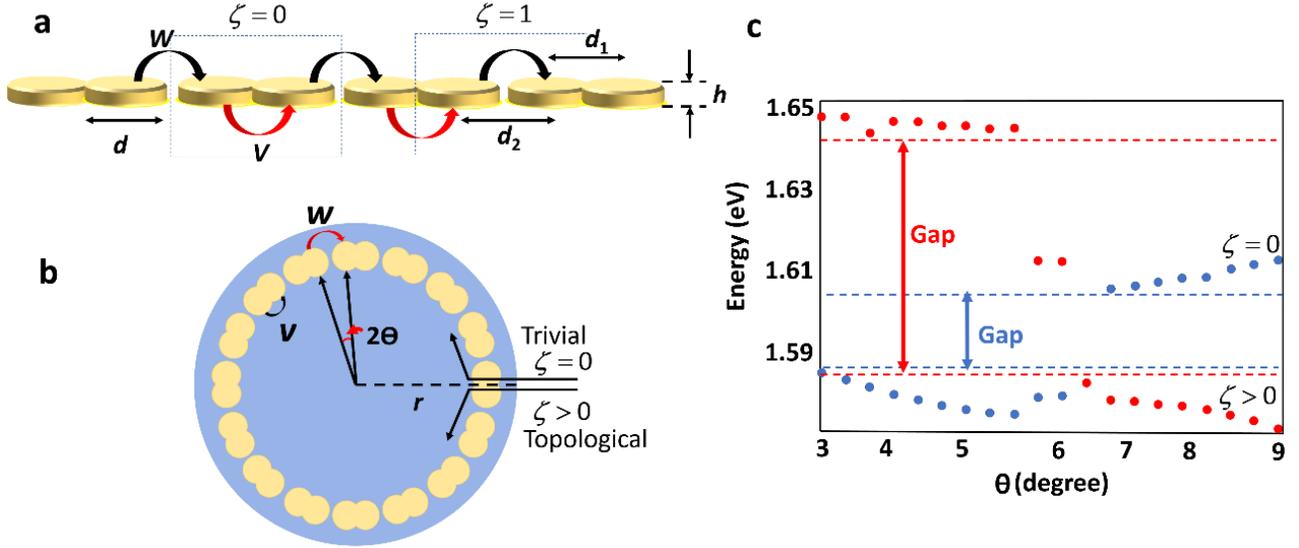

**Figure 1: Schematic and mechanism of a topological ring resonator.** (a) A schematic representation is provided for a one-dimensional plasmonic chain following the Su-Schrieffer-Heeger (SSH) model, which supports localized plasmonic resonances. The model characteristics are determined by the interaction strength between nanoparticles, denoted as $v$ and $w$. Depending on the winding number, $\zeta$, there are two possible choices for the unit cell: $\zeta = 0$ or $1$. (b) A schematic is presented for a plasmonic SSH ring chain structure comprising 16 unit cells. The parameters $d$, $h$, $d_1$, $d_2$, and $r$ respectively represent the diameter, height, gap distance, overlapping distance of the nanoparticles, and the radius of the large disk. The angle between nanoparticles is utilized as the dimerization parameter. A 10 nm-thick $Si_3N_4$ disk serves as the substrate, with gold nanoparticles positioned on top of it. (c) The eigenvalues of the SSH chain in a ring geometry are calculated as a function of the angle between nanoparticles, showcasing the nontrivial and trivial behaviors and highlighting the topological phase transition as the dimerization parameter is varied.

**Topological Phases in the Dipolar Response of Nanoparticles: Tight-Binding Hamiltonian and Theoretical Description**

In this study, we utilize a gold nanoparticle array to construct one-dimensional (1D) plasmonic SSH nanochains. To model the system, we employed the tight-binding (TB) approach, which allows us to simplify the Hamiltonian of the SSH model as follows:[23]

$$\hat{H} = v \sum_{m=1}^{n} \left( |m,B\rangle\langle m,A| + h.c. \right) + w \sum_{m=1}^{n-1} \left( |m+1,A\rangle\langle m,B| + h.c. \right) \quad (3)$$

where $n$ represents the total number of the units of the SSH nanochain, $m$ is the position number, and $v$ and $w$ represent the coupling strengths between intra- and interunits, respectively. Topological edge states only emerge when $v < w$. Therefore, to ensure the presence of topological nontrivial modes, $(w-v)$ should be sufficiently large. For our plasmonic chain, we designed it in a manner where the interactions between gold nanoparticles within the unit are in a strong interaction regime, significantly stronger than the overlapped interactions between neighboring units. To calculate the coupling strength of a gold nanoparticle dimer, we investigated the extinction ratio of an isolated nanoparticle, symmetrically merged, and unmerged gold nanoparticles extending along the direction of a linearly polarized incident light (as depicted in Figure 2).

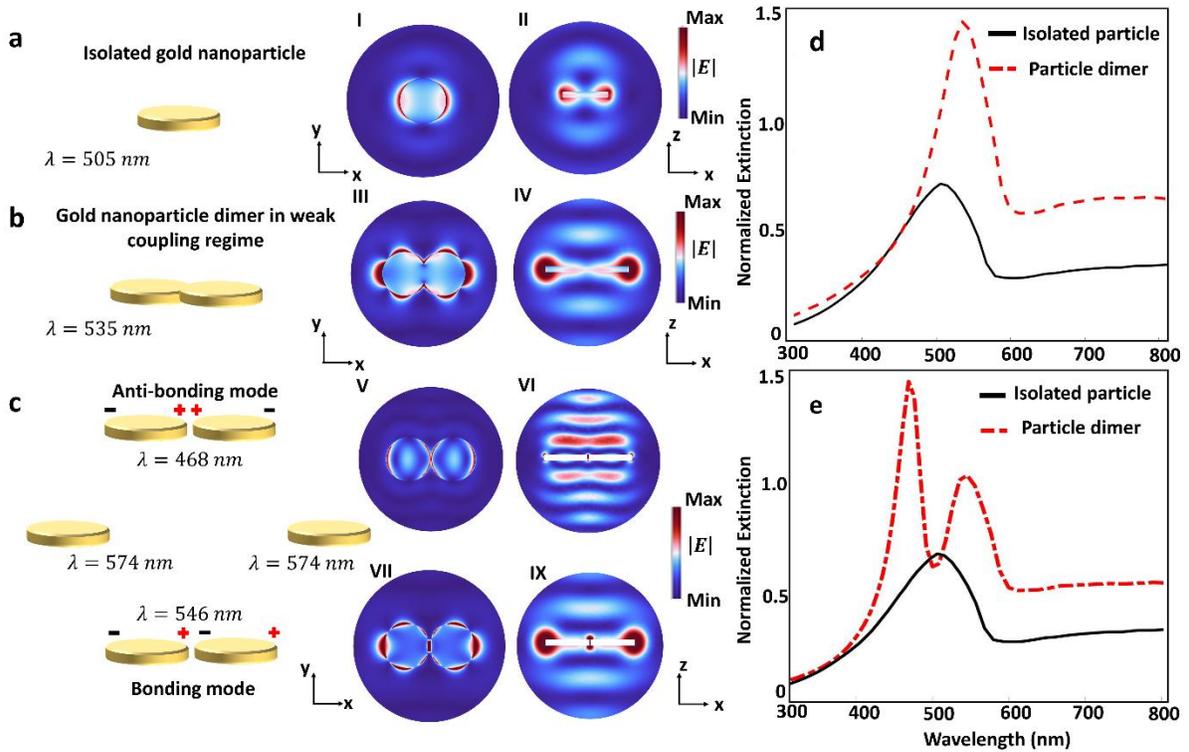

**Figure 2: Plasmon hybridization model.** (a) An isolated plasmonic nanoparticle, where the field distribution in the *xy* (focal plane) (I) and in the *xz* planes (II) are depicted. (b) A symmetric merged gold nanoparticle dimer in the weak-coupling regime. The field distributions in the focal (III) and in the *xz* planes (IV) are shown for the overlapping gold nanoparticles. (c) A symmetric unmerged gold nanoparticle dimer sustaining bonding and antibonding states. The electric field profiles of the plasmonic hybridized modes at the bonding and antibonding eigen energies in the *xy* and *xz* planes (labeled as V-VIII). The extinction spectra for a symmetric merged nanoparticle dimer (d), and a dimer of gold

nanoparticles with a 20 nm gap distance (e), where the spectrum of an isolated nanoparticle is presented for both cases for comparison as well.

In a weak coupling regime, when the diameter $D$ of the gold nanoparticles is given, the coupling strength $g$ between two gold nanoparticles is inversely proportional to the third power of the distance $G$:[24]

$$g = \frac{\omega_0}{2}\left(\frac{D}{2G}\right)^3 \qquad (4)$$

Here, $\omega_0$ represents the energy eigenvalue of an isolated nanoparticle. Therefore, in the weak interaction between two overlapped nanoparticles by a distance ($G$), which does not produce a strong gap field,[20] we can calculate the coupling strength based on the redshift in the resonance frequency of a single nanoparticle, as illustrated in Figure 2(d).

In contrast, when the distance between the nanoparticles is reduced to a very small value, a strong electric field intensity arises in the small gap due to the significant coupling strength between the two nanoparticles. In this scenario, the coupling coefficient deviates from the coupling strength introduced in equation (4). The hybridization model predicts plasmon modes for a symmetric gold nanoparticle dimer oriented along the x-axis, as depicted in Figure 2(a). When the incident beam is polarized along the dimer axis, the in-phase coupling gives rise to a bonding state with a red-shifted resonance, while the out-of-phase coupling results in an antibonding state with a blue-shifted resonance. The situation is reversed for the polarization being perpendicular to the dimer axis (see Supplementary Fig. S2). In this case, the in-phase coupling represents an "antibonding" state with a blue-shifted resonance, and the out-of-phase coupling exhibits a "bonding" state with red-shifted resonance.[26,27] After coupling, the bonding mode becomes lower in energy, whereas the antibonding mode becomes higher in energy. Additionally, the coupling strength is weaker for polarization perpendicular to the dimer axis.

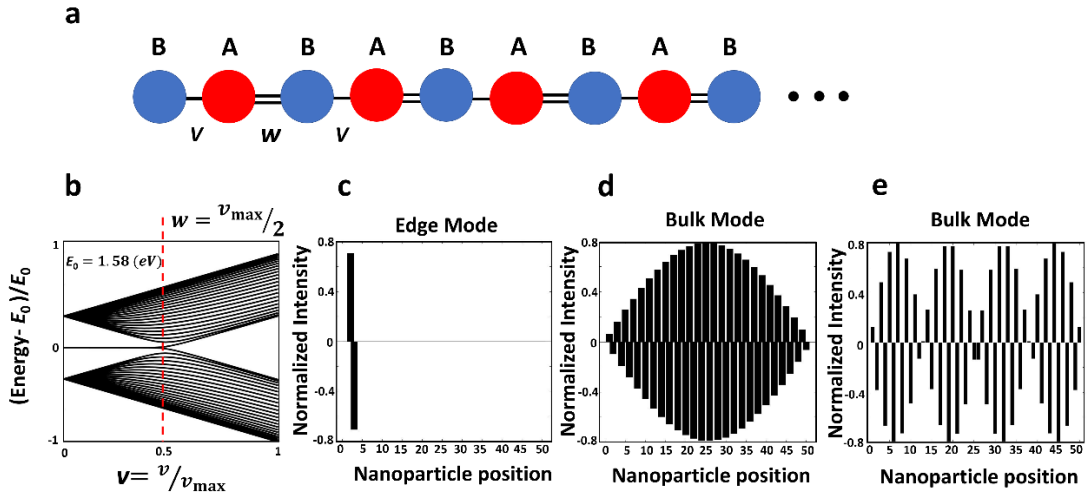

**Figure 3:** (a) The schematic of the plasmonic SSH chain lattice. The parameters are the same as those given in Fig. 1. (b) The energy spectra of the SSH Hamiltonian for a truncated lattice comprising 25 unit cells, versus the coupling constant between the nanoparticles in a plasmonic dimer calculated using a nearest-neighbor tight bonding approach. The horizontal axis is normalized relative to the maximum coupling constant. The energy eigen vectors of the structure calculated at the coupling strength v=0.38, and at the eigen energy of $E = 1.58\ (eV)$ corresponding to the edge mode (c), at $E = 1.48\ (eV)$ (d) and

$E = 1.72\ (eV)$ (e) corresponding to the bulk modes. The nanochain consists of an isolated nanoparticle on the left terminal side so that the total number of the elements is an odd number.

Consequently, plasmon coupling of a symmetric gold nanoparticle dimer can form four states: in-phase bonding, out-of-phase bonding, in-phase antibonding, and out-of-phase antibonding, arranged in order from lower to higher energy (see Section II in the Supporting Information for more details). However, for our investigation of the SSH nanoparticle chain using the tight-binding Hamiltonian, we restrict ourselves to two states: in-phase bonding and out-of-phase antibonding modes. To analyze the interaction Hamiltonian for a two-level system in the strong coupling regime[23] and consider the hybridized bonding and antibonding resonance frequencies obtained from the numerical simulation of the total extinction ratio of a dimer of gold nanoparticles positioned at a 10 nm distance (Figure 2(c)), the coupling strength can be determined by[28] (see Section III in the Supporting Information):

$$\Omega = \omega_+ - \omega_- \qquad (5)$$

where $\omega_+$ and $\omega_-$ are respectively the hybridized antibonding and bonding resonance frequencies. By employing the coupling strengths calculated from equations (4) and (5), we proceed to investigate the energy spectra of the plasmonic nanochain, as illustrated in Figure 1(a). The schematic view of the SSH chain is presented in Figure 3(a), comprising 25 unit cells, where each unit cell consists of two different atoms, represented by red and blue circles. For the sake of simplicity, we label these circles as A and B sites, respectively. The quantum system is simulated within a tight-binding framework, and the energy spectra and energy eigenvectors of the plasmonic SSH chain are obtained by directly diagonalizing its tight-binding Hamiltonian matrix. The coupling between the nearest-neighbor particles, $v$ and $w$, are determined from the localized plasmon modes of gold nanodimers. The energy band structure of the 1D model can be determined by solving for the eigenvalues of the Hamiltonian. In a physical context, when the external degrees of freedom, such as the number of unit cells, are fixed, and a point $k$ in the Brillouin zone and the index $n$ of the energy band are known, it becomes possible to solve for the eigenvalues based on these conditions using Equation (6) (see Section IV in the Supporting Information).[29-30]

$$E(k) = \pm \left| v + ie^{-ik} w \right| = \pm \sqrt{v^2 + w^2 + 2vw \cos k} \qquad (6)$$

From Equation (6), it is evident that when the energy of the system is taken as the vertical coordinate, and $v$ is plotted as the horizontal coordinate while keeping $w$ fixed at half of $v_{max}$, a noticeable occurrence is observed. Specifically, when $v$ is less than $w$, an edge state mode, commonly referred to as the topological state, emerges. To illustrate this, we present the energy eigenvectors at a coupling strength of $v = 0.38 v_{max}$ corresponding to the edge mode with an eigenenergy of $E = 1.58\ (eV)$ and at $v = 0.6 v_{max}$ corresponding to two bulk modes at energies of $E = 1.48\ (eV)$ and $E = 1.72\ (eV)$.

**Numerical results and discussion**

Next, we delve into our numerical results and discussion concerning the plasmonic SSH chain and SSH chain formed into a ring resonator with a symmetry-breaking excitation with electron beams. We explore the critical roles played by the electron impact position in exciting the topological nontrivial or trivial modes.

*Finite Chains and Disorder*

For our investigation, we consider a finite chain and disorder in the system. In the previous section, we studied SSH chain in the quasistatic limit and dipole approximation,[31-32] where the nanoparticles are lossless and chain dimensions are much smaller than the wavelength. However, for equally-spaced chains of nanoparticles with realistic sizes, the quasistatic limit proves insufficient due to the increasing significance of radiative losses for larger particles.[33-34] Therefore, we explore an infinite and truncated chains of gold nanoparticles using full-wave finite-element frequency domain simulations with the COMSOL Multiphysics Software. The photonic band structures, shown in Figure 4(a-c), reveal both topological and trivial photonic band gaps.

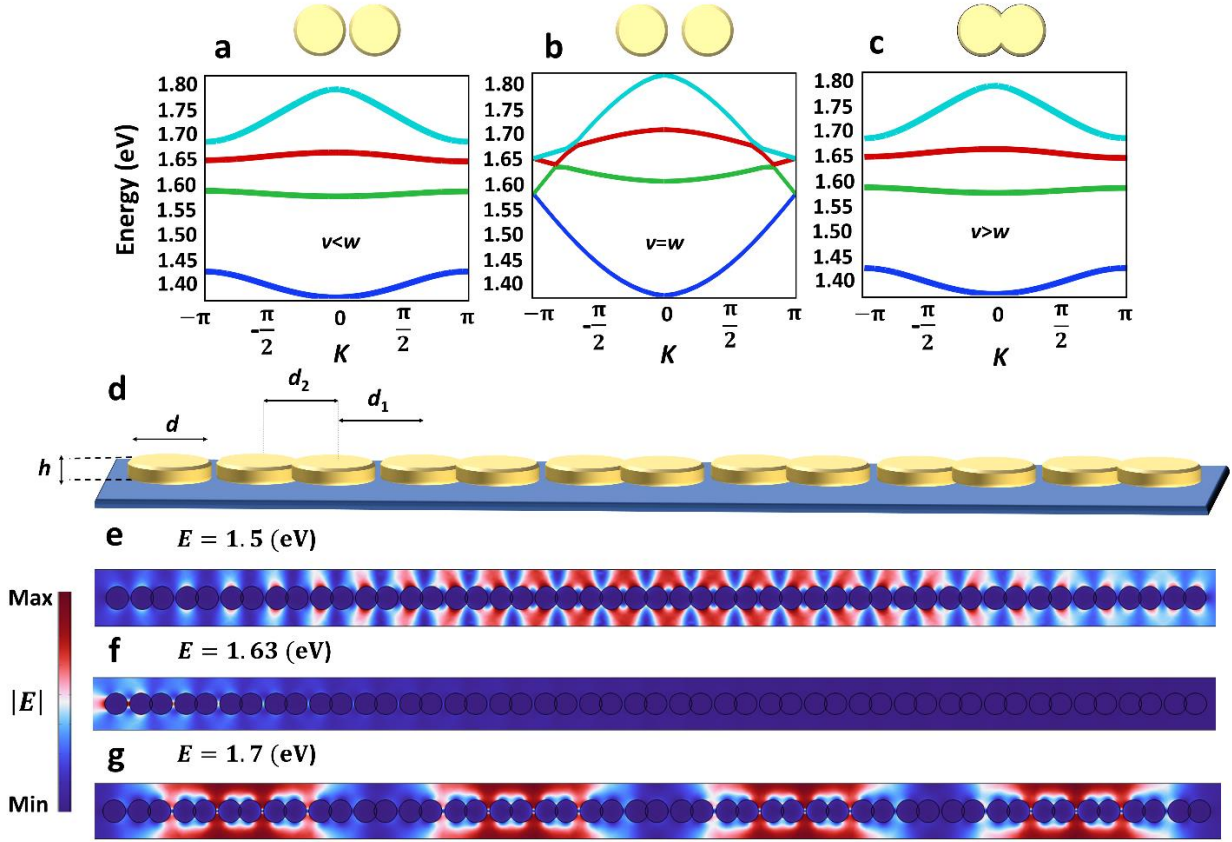

**Figure 4: Numerical simulation findings of the SSH chain composed of gold nanoparticles.** (a-c). The numerically simulated band structures are computed for the nontrivial ($v < w$), an equally spaced ($v = w$), and the trivial lattices ($v > w$) in the momentum space. (d-g). The spatial profile of the electric field shown for the edge mode (f) and two bulk modes (e and g) at depicted energies. for the schematic presented on d for a plasmonic SSH ring chain structure comprising 16 unit cells. The parameters $d$=180 nm, $h$=50 nm, $d_1$=200 nm, and $d_2$=160 nm respectively represent the diameter, height, gap distance, and overlapping distance of the nanoparticles.

By increasing the value of $v$ with respect to $w$, the band structure undergoes a transition from gapped to gapless and again to gapped around the energy of $1.59\ (eV)$, corresponding to changes in the distance between two nanoparticles, indicating a topological phase transition. The distance between the centers of two sublattice particles plays a crucial role in controlling the particle intercell and intracell interactions for the transition from trivial to nontrivial phases. Additionally, we simulate a finite chain of gold nanoparticles with the same parameters. The field distributions ($|E|$) for a truncated structure with $n = 49$ gold nanoparticles, creating a one-dimensional SSH lattice as depicted in Figure 4(d), are shown in Figures 4(e-g). The corresponding eigen energies from top to bottom are $E = 1.59$, $E = 1.63$, and $E = 1.7$ eV, respectively. Notably, the field distribution attributed to the edge mode ($E = 1.63\,(eV)$) is confined to the edge of the array, while the fields of bulk modes are distributed throughout the entire structure. Bulk modes can be identified by their mode profiles, typically resembling normal modes of a chain. The computational outcomes furnish a holistic comprehension of the system shown in Figure 3, that surpasses the confines of the tight-binding computation.

**Su–Schrieffer–Heeger Chain Ring**

In this section, we present our numerical results for the SSH plasmonic ring system under various input conditions and explore the crucial role played by the hopping amplitude on the energy band spectrum and the emergence of topological nontrivial modes. Depending on the loop number ($M$) and the dimerization parameter ($\theta$), different cases can be considered.

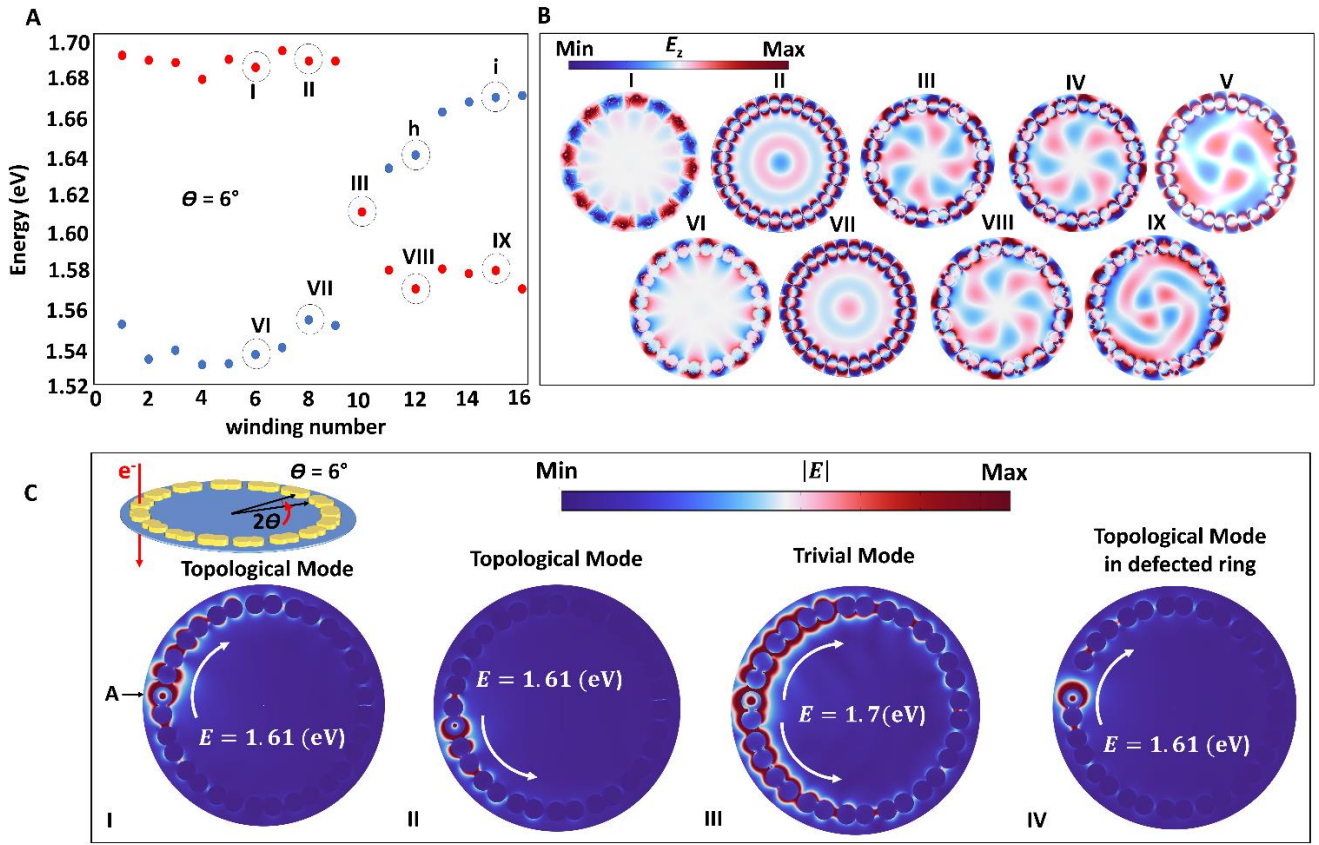

**Figure 5: Symmetry-breaking excitation with electron beams to facilitate the observation of topological edge modes in a plasmonic chain with inherent symmetry.** (a) The eigenvalues of the SSH chain ring resonator for a range of loop numbers (M = 1-16). Topological modes were discerned for M equal to 10, with a dimerization parameter of 6° in a chain comprising 16 unit cells. (b) The spatial profile of the –component of the electric field at a cross-sectional plane for selected modes, corresponding to the points (I to IX) in the eigenvalue diagram. (c) The position of electron impact plays a critical role in

exciting the topological or trivial modes, since it breaks the symmetry of the field distribution. By selectively exciting the A or B sites within the SSH chain possessing a circular geometry, the direction of propagation could be controlled, particularly at the energy associated with the SSH topological mode (I to III). The topological mode is protected by the symmetries and topology of the ring geometry, making it robust enough to withstand local defects caused by the absence of a nanoparticle (IV).

Initially, the eigenvalues of the SSH chain in a ring geometry were calculated as a function of the angle between nanoparticles, revealing a topological phase transition when the dimerization parameter $\theta = 6°$. With $\theta$ fixed at this value, the eigenvalues of the SSH chain ring resonator were computed for a range of loop numbers ($M = 1-16$). Figure 5(a) illustrates the occurrence of topological nontrivial modes for $M = 10$ in a chain comprising 16 unit cells. The topological winding number ($\zeta$) plays a fundamental role in determining the emergent properties of the system, and different homotopy classes with distinct topological properties can be quantified based on this integer value, which counts the number of times the physical space fully covers the order parameter space.[35] The electromagnetic response of the SSH ring structure is directly connected to the winding number (see Section V in the Supporting Information for more details of TB in a ring).

To understand this relation, we need to start from the pointing vector. The pointing vector of the electromagnetic field, indicating the direction and magnitude of the field power ($\vec{P}$). The polarization of light arises from the intrinsic spin nature of light, resulting in an average spin angular momentum of zero for the resonant mode circulating in the ring resonator. On the other hand, the orbital angular momentum is a consequence of the spatial structure imposed on the shape of the wavefront, characterized by the phase factor $e^{i\zeta\varphi}$ ($\varphi$ represents azimuthal angle). This phase becomes undefined on the beam z-axis, and $\zeta$ is the OAM component along the quantization axis of z.[36] Applying the angular momentum operator to the propagation axis, $\hat{L}_z = i\hbar \frac{\partial}{\partial \varphi}$, it is clear that the discrete integer value $\zeta$, often referred to as the topological charge, determines the quantized angular momentum eigenvalue: $\zeta = 2M \, (for \, M = \pm 1, \pm 2, \pm 3,...)$, where $\zeta$ is the winding number, and the phase $e^{i\zeta\varphi}$ winds around the origin $\zeta$ times as the spatial azimuthal angle $\varphi$ circles the z-axis once. Another important point here is, since there is circular symmetry in the ring resonator, both positive and negative winding number should be present for a specific topological SSH ring mode. The emergence of two distinct topological modes within the revealed bandgap, as depicted in Figure 1, can be attributed to the excitation of two counter-rotating circular modes. The manifestation of positive and negative winding numbers becomes apparent upon analyzing the orbital angular momentum associated with these modes. Detailed visualizations of the orbital angular momentum pertaining to both of these topologically nontrivial modes are presented in Figure S4 (see Section VI in the Supporting Information). In our investigation, we have examined the spatial profile of the z-component of the electric field at a cross-sectional plane for specific modes, corresponding to points (I to IX) in the eigenvalue diagram. These modes are characterized by distinct loop numbers, namely 6, 8, 10, 12, and 15 (Figure 5(b)).

Remarkably, an intriguing phenomenon emerges when we traverse the structure and encircle it. The direction of our walk, either counterclockwise or clockwise, plays a pivotal role in sensing the topological nature of the chain. For instance, when we walk counterclockwise around the structure starting from a particular position shown as point A in Figure 5(c) we perceive a topological nontrivial chain. Conversely, walking clockwise from the same starting position leads us to sense a topological trivial chain. To explore this phenomenon in the plasmonic SSH ring we employed symmetry-breaking excitation with electron beams. Given the pivotal role played by the electron impact position in the excitation of topological or trivial modes, resulting from the symmetry disruption in the field distribution, a judicious selection of the A or B sites within the circularly arranged SSH chain holds the potential to govern the propagation direction. This control is particularly pertinent to energy levels associated with the SSH topological mode, as elucidated in Figures 5(c) (I to III). The inherent resilience of the topological mode, hinged upon the symmetries and topology inherent to the ring geometry, is vividly showcased in its capacity to endure against local defects, such as the absence of a nanoparticle, as depicted in Figure 5(c) (IV).

**Conclusion**

In conclusion, our investigation into topological plasmonic chains and quantum rings has unveiled a rich tapestry of phenomena with profound implications for nanophotonics and quantum technologies. Through careful analysis of the one-dimensional Su-Schrieffer-Heeger (SSH) model adapted to metallic nanoparticle arrays, we have demonstrated the emergence of topological nontrivial modes, characterized by protected edge states. By exploring the intricate interplay between nanoparticle interactions, dimerization, and periodicity, we have unraveled the critical parameter regimes that lead to the formation of topologically distinct phases. The topological invariant winding number, has served as a powerful tool to identify and characterize these phases, providing a deeper understanding of the topological behavior exhibited by our plasmonic systems. Our thorough numerical analysis of finite chains and disorder has extended the exploration beyond the quasistatic limit, accounting for the impact of radiative losses. This comprehensive approach has enhanced the relevance and applicability of our findings to experimental settings, particularly in plasmonic metamaterials. Of particular significance is our investigation into the SSH ring system, which has shed light on the fascinating connection between the electromagnetic response and the winding number. Moreover, our exploration has been enriched by the use of symmetry-breaking electron beam excitation, providing valuable insights into the directional control of plasmonic modes, particularly within the topological regime. This knowledge holds immense promise for advanced applications in quantum information processing and sensing.

**Supporting Information**

The Supporting Information is available free of charge at https://pubs.acs.org.

Detailed information on topological ring field profile in both trivial and nontrivial modes, the interaction Hamiltonian in a two-level system, the plasmon hybridization mode in gold nanodisk dimers, the tight bonding Hamiltonian in plasmonic chain and ring, more details on visualization of the orbital angular momentum and winding number in the topological ring, and theoretical and simulation methods.

**Corresponding Author**

Fatemeh Davoodi- Institute for Experimental and Applied Physics, Kiel University, 24118 Kiel, Germany; Email: davoodi@physick.uni-kiel.de.

Nahid Talebi- Institute for Experimental and Applied Physics, Kiel University, 24118 Kiel, Germany; Email: Talebi@physick.uni-kiel.de.

# Topological Plasmonic Ring Resonator

*Fatemeh Davoodi[1],\* and Nahid Talebi[1,2],\**

[1]*Institute of Experimental and Applied Physics, Kiel University, 24098 Kiel, Germany*

[2]*Kiel Nano, Surface and Interface Science KiNSIS, Christian Albrechts University, Kiel, Germany*

In this Supporting Information, we provide further details on topological ring field profile in both trivial and nontrivial modes, the interaction Hamiltonian in a two-level system, the plasmon hybridization mode in gold nanodisk dimers, the tight bonding Hamiltonian in plasmonic chain and ring, more details on visualization of the orbital angular momentum and winding number in the topological ring, and simulation methods:

I. Topological SSH ring field profile
II. The interaction Hamiltonian for a two-level system
III. Plasmon hybridization in individual gold nanodisk dimers
IV. Tight bonding Hamiltonian in SSH plasmonic chain
V. Tight bonding Hamiltonian in ring
VI. Visualization of the orbital angular momentum for both topologically nontrivial modes
VII. Methods

## Section I- Topological SSH ring field profile

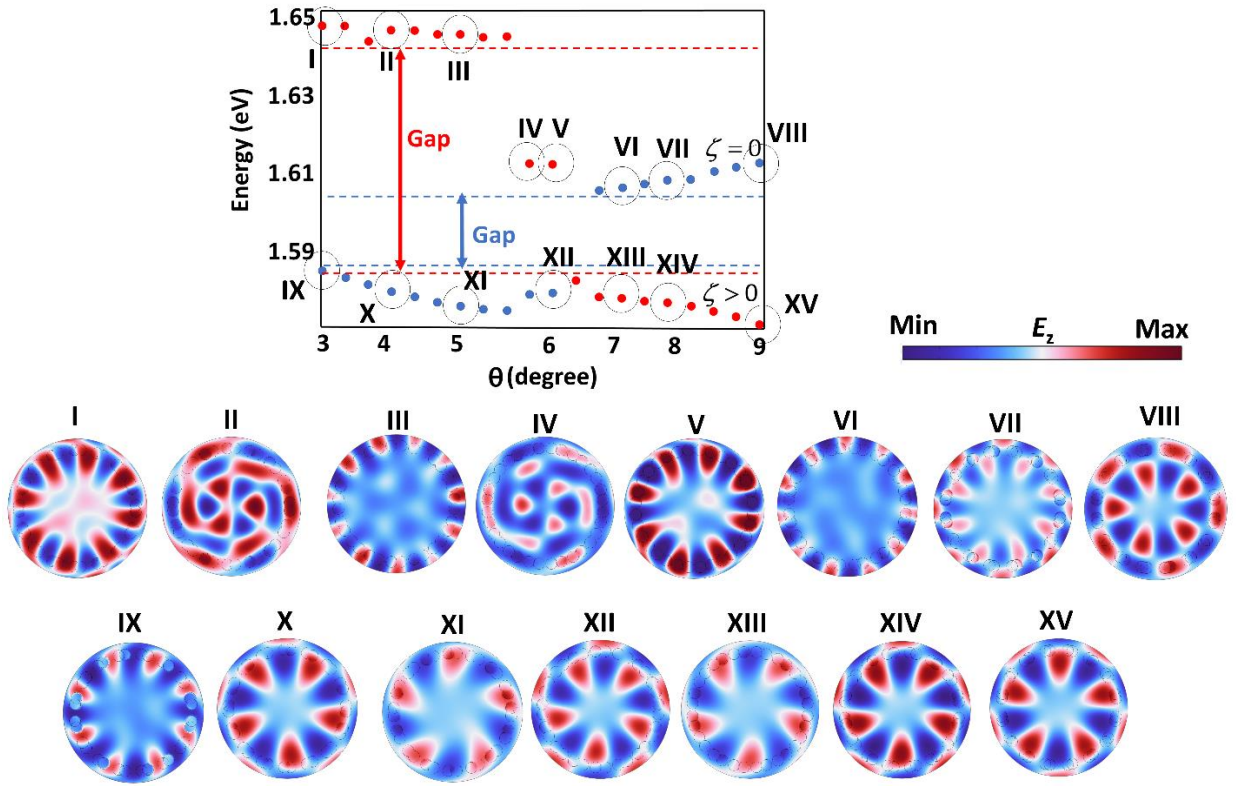

**Figure S1:** The spatial profile of the z–component of the electric field at a cross-sectional plane for selected modes, corresponding to the points (I to XV) in the eigenvalue diagram.

## Section II- Plasmon hybridization in individual gold nanodisk dimers

The plasmon modes of a symmetric gold nanoparticle dimer aligned along the x-axis were predicted using the hybridization model, as illustrated in Figure S2. When an incident beam is polarized along the dimer axis, the in-phase coupling results in a bonded state exhibiting a resonance with a redshift, while the out-of-phase coupling leads to an antibonding state with a resonance blue shift. Conversely, for polarization perpendicular to the dimer axis, the in-phase coupling generates an antibonding state with a blue-shifted resonance, whereas the out-of-phase coupling manifests as a bonded state with a resonance that is red-shifted. Following the coupling process, the energy of the bonded mode decreases, while the energy of the antibonding mode increases. Notably, the strength of coupling is attenuated for polarization perpendicular to the dimer axis.

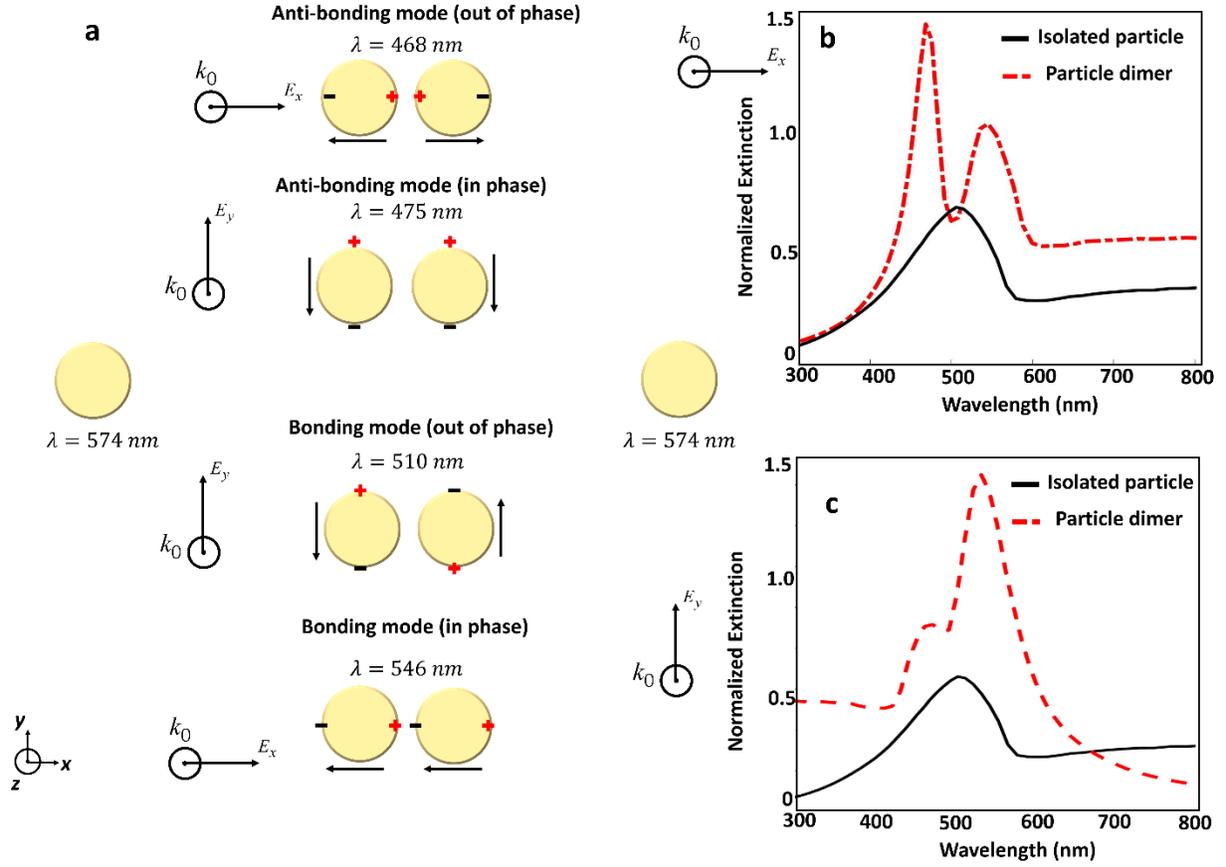

**Figure S2: Plasmon hybridization model in different excitation.** (a) A symmetric unmerged gold nanoparticle dimer sustaining four bonding and antibonding states. (b) The extinction spectra for a symmetric merged nanoparticle dimer in x-polarized Excitation (c) The extinction spectra for a symmetric merged nanoparticle dimer in y-polarized Excitation.

### Section III- The interaction Hamiltonian for a two-level system

Every individual nanoparticle may be conceptualized as a singular oscillator, with the interaction amongst gold nanoparticles being expounded upon as a two-level system engaged in mutual interaction. By employing boson, $\hat{a}, \hat{a}^\dagger$, operators to quantize the plasmon field, the description of the system can be formulated using the Jaynes-Cummings Hamiltonian as outlined in prior research: [1,2]

$$\hat{H} = \hbar\omega_{pl,1}\,\hat{a}^\dagger\hat{a} + \hbar\omega_{pl,2}\,\hat{a}^\dagger\hat{a} + \hbar g\left(\hat{a}^\dagger\hat{\sigma} + \hat{\sigma}^\dagger\hat{a}\right) \qquad (1)$$

where $\hat{\sigma}$ is the Pauli raising operator for the two-level system, $\omega_{pl,1}$ and $\omega_{pl,2}$ are the plasmon frequency 1 and 2, $g$ is the coupling strength between two localized plasmons. $\hbar g\left(\hat{a}^\dagger\hat{\sigma} + \hat{\sigma}^\dagger\hat{a}\right)$ term can be considered as the interaction Hamiltonian and can be described in the nanoparticle interaction as

$$H_{\mathrm{int}} = -\frac{1}{2}\alpha(\omega)E_1 E_2 \quad (2)$$

$\alpha(\omega)$ is the polarizability of the nanoparticles. $E_1$ and $E_2$ are the electric field vectors associated with the oscillations of the nanoparticles. In the quasistatic limit and with neglecting radiative damping, the polarizability, the electric field generated by a point dipole $p$ oscillating with frequency $\omega$:[3]

$$E(r,t) = \left[\left(1 - \frac{i\omega r}{c}\right)\frac{3\hat{r}.p\hat{r} - p}{r^3} + \frac{\omega^2}{c^2}\frac{p - \hat{r}.p\hat{r}}{r}\right]e^{i\omega r/c} \quad (3)$$

Here $r$ is the position vector. In the dimmer case spaced a distance $d$, in the absence of an applied field, the field at each dipole is the sum of the fields due to all the other dipole. The induced moment of the dimmer is the polarizability $\alpha(\omega)$ times this field[3]:

$$p_n = \alpha(\omega) \sum_{m \neq n} \left(\left(1 - \frac{i\omega|n-m|d}{c}\right)\frac{3\hat{r}.p_m\hat{r} - p_m}{|n-m|^3 d^3} + \frac{\omega^2}{c^2}\frac{p_m - \hat{r}.p_m\hat{r}}{|n-m|d}\right)e^{i\omega|n-m|d/c} \quad (4)$$

In the case of a thin circular disk with a radius "$a$" and the disk thickness much smaller than the radius, the polarizability of can be approximated as[4]

$$\alpha(\omega) = \frac{\varepsilon_0 \pi a^3}{\varepsilon(\omega) - \varepsilon_0} \quad (5)$$

$\varepsilon(\omega)$ is the absolute permittivity (dielectric constant) of the disk material, $\varepsilon_0$ is the absolute permittivity of vacuum.

Diagonalization of the Hamiltonian Eq. 1, the eigen frequency of the system after interaction gives[5]:

$$\omega_\pm = \frac{1}{2}\left(\omega_{pl,1} + \omega_{pl,2} - i(\gamma_1 + \gamma_2)\right) \pm \Omega_R \quad (6)$$

$\Omega_R = \sqrt{g^2 + \frac{1}{4}(\omega_{pl,1} - \omega_{pl,2})^2}$ is the Rabi frequency, and $\gamma_1$ and $\gamma_2$ the damping of oscillators. Since our system is symmetric $\omega_{pl,1} = \omega_{pl,2}$ and $\gamma_1 = \gamma_2$. In this particular scenario, the coupling constant coincides with the Rabi frequency. Through the process of data fitting utilizing dimer simulations, we were able to determine the relationship between the coupling constant and the separation distance of nanoparticles within gold nanoparticle dimers.

## Section IV- Tight bonding Hamiltonian in SSH plasmonic chain

To model the system, we employed the tight-binding (TB) approach, which allowed us to simplify the Hamiltonian of the SSH model as follows: [6,7]

$$\hat{H} = v \sum_{m=1}^{n} \left( |m,B\rangle\langle m,A| + h.c. \right) + w \sum_{m=1}^{n-1} \left( |m+1,A\rangle\langle m,B| + h.c. \right) \quad (7)$$

$$|m,\alpha\rangle = |m\rangle \otimes |\alpha\rangle, \; m = 1,...,N, \alpha = A, B$$

$$\hat{H} = v \sum_{m=1}^{n} \left( |m,B\rangle\langle m| \otimes \sigma_x + h.c. \right) + w \sum_{m=1}^{n-1} \left( |m+1\rangle\langle m| \otimes \frac{\sigma_x + i\sigma_y}{2} + h.c. \right) \quad (8)$$

Looking for the eigenstates: $\hat{H}_{bulk} |\psi_n\rangle = E_n |\psi_n\rangle, n = 1,....2N$

By using plane wave solutions:

$$|k\rangle = \frac{1}{\sqrt{N}} \sum_{m=1}^{N} e^{imk} |m\rangle, \; k \in \left\{ \frac{2\pi}{N}, \frac{4\pi}{N},..., \frac{2N\pi}{N} \right\}$$

Total eigenstates: $|\psi_n(k)\rangle = |k\rangle \otimes |u_n(k)\rangle$, and

$$|u_n(k)\rangle = a_n(k)|A\rangle + b_n(k)|B\rangle, \; n = 1,2$$

The vectors $|u_n(k)\rangle$ are the eigenstates of the bulk momentum-space Hamiltonian: [8]

$$H(k)|u_n(k)\rangle = E_n(k)|u_n(k)\rangle \quad (9)$$

$$H(k) = \begin{pmatrix} 0 & v + we^{ik} \\ v + we^{ik} & 0 \end{pmatrix} \Rightarrow E_{\pm}(k) = \pm |v + we^{ik}| = \pm\sqrt{v^2 + w^2 + 2vw\cos(k)} \quad (10)$$

## Section V- Tight bonding Hamiltonian in ring

The image in Figure S3 depicts the ring structure used to examine the eigenstate of the SSH ring. The ring comprises N unit cells, and each unit cell consists of two distinct atoms denoted by red and blue disks. We refer to these disks as A and B sites, respectively, for simplicity. The quantum system is simulated using a tight-binding framework.

$$H = \sum_n v \left( e^{i\theta} a^{\dagger}_{\alpha,n} a_{\beta,n} + e^{-i\theta} a^{\dagger}_{\beta,n} a_{\alpha,n} \right) + \sum_n w \left( e^{i\theta} a^{\dagger}_{\beta,n} a_{\alpha,n+1} + e^{-i\theta} a^{\dagger}_{\alpha,n+1} a_{\beta,n} \right) \quad (11)$$

where $n$ is the unit cell index and it runs from 1 to N, $a^{\dagger}_{\alpha(\beta),n}$ and $a_{\alpha(\beta),n}$ are the bosonic operators. $v$ and $w$ represent the intra- and inter-cell hopping coefficients. Phase factor $\theta$ is introduced in the hopping term called AB phase which is expressed as $\theta = \frac{2\pi}{N}$, $N$ is the total number of sites in the ring, the site number with $N+1 = 1$.

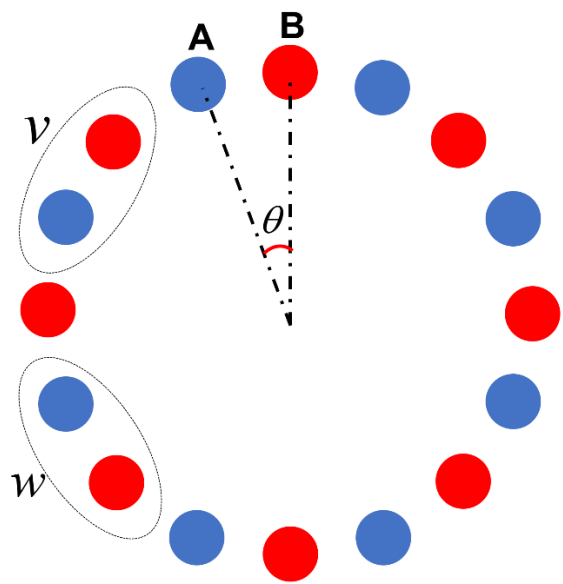

**Figure S3:** SSH ring chain structure comprising 16 sites. The angle between nanoparticles is utilized as the dimerization parameter.

## Section VI- Visualization of the orbital angular momentum for both topologically nontrivial modes

The Poynting current can be written as the some of spin and orbital flow densities:[9]

$$P = P_{orb} + P_{spin} \qquad (12)$$

$$P_s = \frac{-c\varepsilon_0}{8\omega} \text{Im}\left[\nabla \times \left(E^* \times E + H^* \times H\right)\right] \qquad (13)$$

$$P_o = \frac{c\varepsilon_0}{4\omega} \text{Im}\left[\left(E^*.\nabla\right)E + \left(H^*.\nabla\right)H\right] \qquad (14)$$

To Visualize the rotational behavior of angular momentum and calculation of winding number, we computed the orbital angular momentum vorticity:[10]

$$\Omega_o = \nabla \times P_o = \frac{c\varepsilon_0}{4\omega} \text{Im}\left[\nabla E^*.(\times\nabla)E + \nabla H^*.(\times\nabla)H\right] \qquad (15)$$

where $\nabla E^*.(\times\nabla)E = \nabla E_x^* \times \nabla E_x + \nabla E_y^* \times \nabla E_y + \nabla E_z^* \times \nabla E_z$

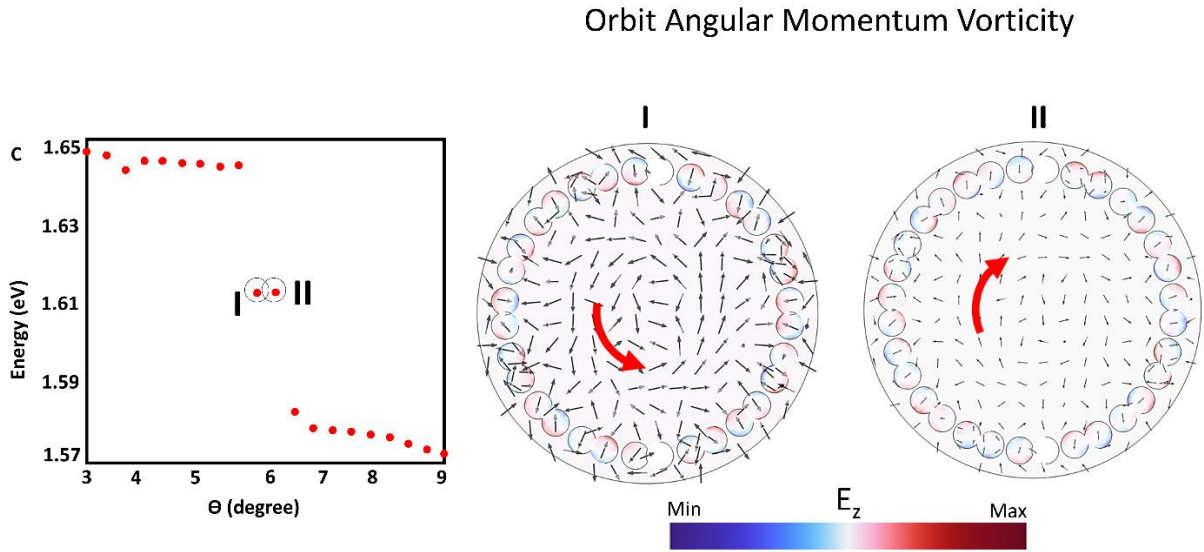

**Figure S4:** Positive and negative winding numbers becomes apparent upon analyzing the orbital angular momentum vorticity associated with the points I and II in the eigenvalue diagram to show the direction of rotation in orbital angular momentum. The spatial distribution of the z-component of the electric field is depicted at the surface of the plasmonic structure.

## Section VI- Methods

We have employed both a home-built analytical solution code in tight-binding (TB) approach and the COMSOL Multiphysics software to gain insight into eigen states of SSH chain, SSH ring, the eigenvalues of the plasmonic SSH chain in a ring geometry, and using symmetry-breaking excitation technique based on electron beams for excitation and control the unidirectional propagation in SSH plasmonic ring resonator. As discussed in Supplementary sections III-V, we employed tight bonding Hamiltonian for calculation of eigen states of SSH plasmonic chain. To calculate the intercell and intracell coupling strengths of two gold nanodisk dimmer, we used numerical simulation Wave Optics toolbox of COMSOL in a 3D simulation domain, which is based on solving the Maxwell equations in real space and in the frequency domain. For symmetry breaking excitation, we model an electron beam by a current density distribution corresponding to a swift electron at a kinetic energy of $U = 30\,keV$. We have utilized an oscillating "edge current" as an electron beam along a straight line representing the electron beam. The current was expressed by $I = I_0 \exp(i\omega z/v_e)$. We apply the free tetrahedral mesh with refined elements close to the electron's trajectory, it is 0.04 nm along the electron trajectory. We allow for an increase of the size of the mesh elements towards outer boundaries of the simulation domain. The area of PML is meshed by 25-layer Swept meshing. Because the Geometry of the multilayered structure, Perfectly Matched Layers (PML) help to attenuate the electric field at the boundaries of the simulation domain and prevent unphysical field reflections from the boundaries.